# Modeling Three and Four Coupled Phase Qubits

Zechariah E. Thrailkill, S. T. Kennerly, and R. C. Ramos

*Abstract*—The Josephson junction phase qubit has been shown to be a viable candidate for quantum computation. In recent years, the two coupled phase system has been extensively studied theoretically and experimentally. We have analyzed the quantum behavior of three and four capacitively-coupled phase qubits with different possible configurations, using a two-level system model. Energy levels and eigenstates have been calculated as a function of bias current and detuning. The properties of these simple networks are discussed.

## I. INTRODUCTION

ONE of the key requirements towards building a superconductor-based quantum computer is the coupling of Josephson junction qubits. The quantum entanglement that results from coupling multiple qubits is essential in implementing important applications such as quantum state transfer and error correction. This has motivated many theoretical and experimental studies on two coupled superconducting charge qubits [1]-[2], phase qubits [3]-[6] and flux qubits [7]-[8]. There have also been spectroscopic studies on three- and four-coupled flux qubits [9]-[10], multiple-particle entangled states [11] and two phase qubits coupled to a resonant cavity [12].

We present theoretical simulations of systems involving three and four phase qubits capacitively-coupled together in different configurations. Phase qubits are compact, tunable devices and are one of the strongest candidates for quantum computing. Recently, the issue of how multiple phase qubits can be networked together and what entangling protocols can correspondingly be implemented have received a significant amount of attention [13]-[14]. In order for phase qubits to be used in quantum computing, the properties of larger systems of coupled qubits must be analyzed. In this paper, we calculate the energy level spectra, describe entangled states and the properties that arise from the simple networks of three and four coupled phase qubits.

## II. THEORY

We consider a network of identical Josephson phase qubits that are coupled to each other with identical capacitors and arranged in various configurations. The Hamiltonian for such a general system can be written as

$$H = \frac{1}{2}\left(\frac{\Phi_0}{2\pi}\right)^{-2} [p_1, p_2, \cdots] \begin{bmatrix} C \end{bmatrix}^{-1} \begin{bmatrix} p_1 \\ p_2 \\ \vdots \end{bmatrix} + W(\gamma_1) + W(\gamma_2) + \cdots \quad (1)$$

where each junction has an intrinsic capacitance $C_J$ and critical current $I_c$. Here $\Phi_0 = h/(2e) = 2.07 \times 10^{-15}$ Weber is the fundamental flux quantum. The variables for junction $i$ are the phase difference $\gamma_i$ across the junction and the conjugate momentum $p_i$, while the potential energy $W(\gamma_i)$ contains the washboard potential for each junction $i$. In general, the conjugate momenta of different junctions are coupled to each other by the inverse of the capacitance matrix $C$. This capacitance matrix is determined by the intrinsic capacitance $C_J$, the coupling capacitance $C_c$ and the particular configuration of the qubits as represented by the matrix elements $C_{ij} = C_J\delta_{ij} + C_c\mathcal{L}_{ij}$. Here $\mathcal{L}_{ij}$ is the circuit's Laplacian matrix, which is found by reversing the sign of the adjacency matrix $A_{ij}$ and replacing each diagonal element $A_{ii}$ with the degree (number of connections) of vertex $i$ [16]. In graph theory, the adjacency matrix $A_{ij}$ of a simple graph is formed by writing "1" for element $A_{ij}$ if vertex $i$ is connected to vertex $j$

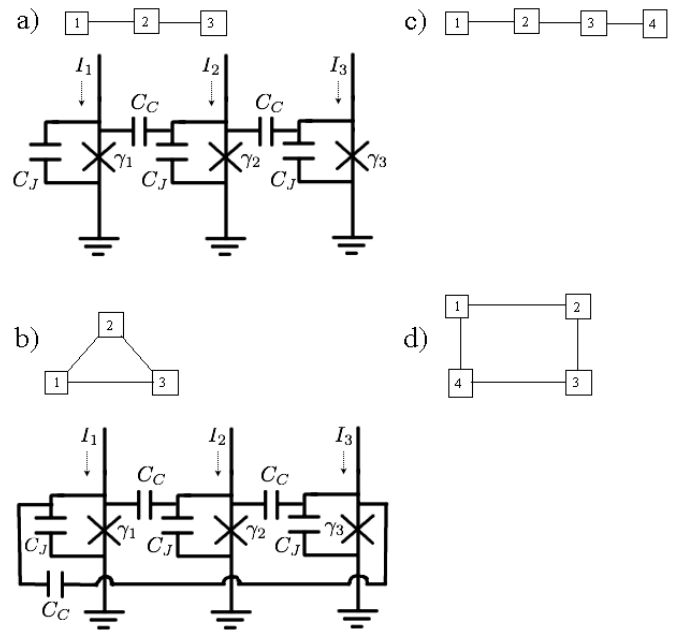

Fig. 1. (a) The linear three-qubit system and its corresponding circuit diagram. Each numbered square represents a junction and each line connecting them represents a coupling capacitor. (b) The triangular three-

qubit system and its circuit diagram. (c) The linear four-qubit system. (d) The four-qubit box system.

and "0" otherwise. By approximating the qubits as two level systems and limiting the coupled system to energy states with one excitation, the time-independent Hamiltonian can be written as

$$H \approx \sum_{i \leq j} \hbar \sqrt{\omega_i \omega_j} [M]_{ij} \left( \hat{s}_i \hat{s}_j^\dagger + \hat{s}_i^\dagger \hat{s}_j \right) \quad (2)$$

where $\omega_i^2 = \frac{\Phi_0 I_c}{2\pi C_J} \sqrt{1 - J_i^2}$ (3)

Here, $\hat{s}^\dagger$ and $\hat{s}$ are the raising and lowering operators for the two level system, $M_{ij} = C_J\, C_{ij}^{-1}$ is the unit-less coupling matrix, $\omega_i$ is the plasma frequency for qubit $i$ and $J_i = I_b/I_c$ is the reduced bias current for qubit $i$. The coupling matrix for the linear three qubit system in Fig. 1a is found to be:

$$M_{ij} = \frac{1}{1+2\kappa} \begin{bmatrix} 1+\kappa-\kappa^2 & \kappa & \kappa^2 \\ \kappa & 1 & \kappa \\ \kappa^2 & \kappa & 1+\kappa-\kappa^2 \end{bmatrix} \quad (4)$$

where $\kappa = C_c/(C_J + C_c)$ is the coupling strength. The coupling matrix for the triangular three qubit system in Fig. 1b is:

$$M_{ij} = \frac{1}{1+2\kappa} \begin{bmatrix} 1 & \kappa & \kappa \\ \kappa & 1 & \kappa \\ \kappa & \kappa & 1 \end{bmatrix} \quad (5)$$

Comparing (4) and (5), it is clear that the linear system's coupling matrix element $M_{22}$, corresponding to qubit 2, is different from the diagonal terms for qubits 1 and 3, $M_{11}$ and $M_{33}$. This is because qubit 2 is connected to two qubits while qubits 1 and 3 are connected to only one. It is important to note that the conjugate momenta in (1) are proportional to the charge on the junction and the coupling capacitors connected to it, not just the charge on the junction. The coupling capacitors have the effect of lowering the energy levels of a qubit. There is also a difference in the off-diagonal terms, $M_{13}$ and $M_{31}$, that couple qubits 1 and 3. In the linear system, this is expected since the coupling between non-adjacent qubits is second order, and thus much weaker, while in the triangular system there is a connection between all qubits.

## III. RESULTS OF SIMULATIONS

In order to utilize these systems in quantum computation, the properties of the various qubit configurations must be examined. One such property is perfect state transfer. That is, if one qubit is prepared in a superposition of its ground and first excited state, $\alpha|0\rangle+\beta|1\rangle$, while a target qubit and all other qubits are in their ground state, then after a period of time passes the system will evolve to where the target qubit is in the $\alpha|0\rangle+\beta|1\rangle$ state and the others are in their ground state [13]. This will be discussed for the different qubit configurations. The energy spectrum and eigenstates were calculated using the Jacobi transformation method for matrix diagonalization for three and four qubit systems in various geometries. In these simulations, we used a characteristic junction plasma frequency $\omega/2\pi = 5.5$ GHz with a reduced bias current $J = 0.985$ and a coupling strength $\kappa = 0.01$.

### A. Linear System

We first consider three qubits in a linear chain as shown in Fig. 1a, which is an extension of the well-studied two qubit system [3]-[6]. For concreteness, we fix the bias currents $J_2 = 0.979$ and $J_3 = 0.985$ while ramping the bias current $J_1$ of qubit 1 through these values. Since qubits 2 and 3 are detuned from each other, ramping the current $J_1$ is expected to entangle qubit 1 with qubits 2 and 3, separately. This is consistent with the results of our simulations, as shown in Fig. 2 where the calculated energy level spectrum shows two avoided crossings close to the fixed values of $J_2$ and $J_3$. Strictly speaking, the point where qubits 1 and 2 become strongly entangled is not exactly where these currents are equal, but rather at a point where $J_2$ is slightly less than $J_1$. This is because qubit 2 has one more coupling capacitor than qubits 1 and 3.

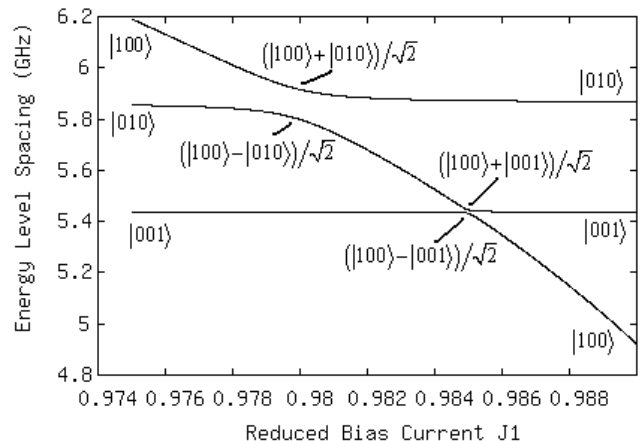

Fig. 2. Energy level spectrum for the linear three qubit system. Here $J_2 = 0.979$, $J_3 = 0.985$, while $J_1$ is being ramped. Qubits 1 and 2 become entangled when $J_1$ is slightly detuned from $J_2$ but far from $J_3$. Qubits 1 and 3 become entangled when $J_1 = J_3$. The energy gaps of the avoided crossings are different because qubits 1 and 3 have much weaker coupling than 1 and 2.

On the other hand, when $J_1 = J_3$ then qubits 1 and 3 become maximally entangled. Since both qubits are on the ends of the linear system and have the same number of coupling capacitors, their corresponding terms in the coupling matrix are equal and the two qubits become entangled when their bias currents are equal. Furthermore, the energy gap of this avoided crossing is much smaller than that between the entangled states of 1 and 2. This is because the coupling term between qubits 1 and 3 is of second order. One consequence of the small energy gap is that, while state transfer is possible between the two end qubits, it would be a factor of $1/\kappa$ slower than that between adjacent qubits.

When all three bias currents are equal, $J_1 = J_2 = J_3$, the entire system is in an entangled state, with eigenstates shown in Fig. 3. When qubit 1 is detuned from qubits 2 and 3, the eigenstates of the isolated two-qubit system do not form symmetric and anti-symmetric eigenstates. This is shown by the particular coefficients appearing on the left side of Fig. 3. This asymmetry results from unequal diagonal terms in the coupling matrix (4), i.e. $M_{22} \neq M_{33}$. This inequality can impact how the system behaves. For instance, even with qubit 1 greatly detuned, total state transfer is not possible between qubits 2 and 3.

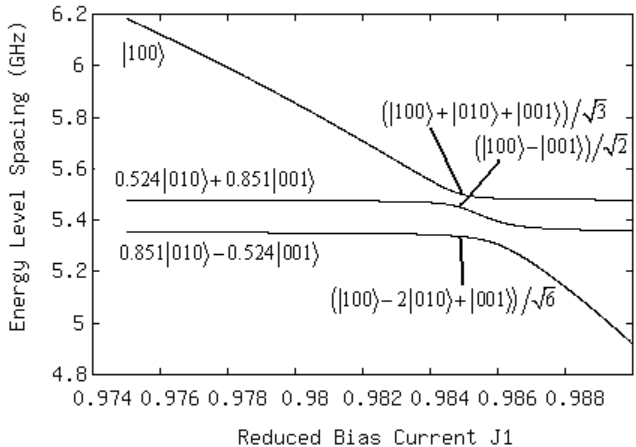

Fig. 3. Energy level spectrum for the linear three qubit system. Here $J_2$ = 0.985 and $J_3$ = 0.985 while $J_1$ is being ramped. On the right are the eigenstates of the system when all the bias currents are equal. On the left are the eigenstates when $J_1$ is far from $J_2$ and $J_3$, where qubits 2 and 3 form an asymmetric superposition of their eigenstates.

The symmetry in these states can be restored by changing the bias current $J_2$ to the value

$$J_2^2 = 1 - (1+\kappa)^4 (1 - J_3^2). \tag{6}$$

This compensates for the decreased energy level of qubit 2 caused by its extra coupling capacitor. The resulting eigenstates become symmetric and anti-symmetric about qubits 2 and 3, as shown in Fig. 4. This allows for perfect state transfer between qubits 2 and 3 [16]. Also, when $J_1 = J_3$ and $J_2$ takes on the value in (6), the eigenstates become symmetric and anti-symmetric about qubits 1 and 3. This system then allows for state transfer between qubits 1 and 3. When all the qubits are entangled the anti-symmetric state is the second energy level as seen in Fig. 4. The symmetric state is an equal superposition of the other two states. State transfer between qubits 1 and 3 is then possible.

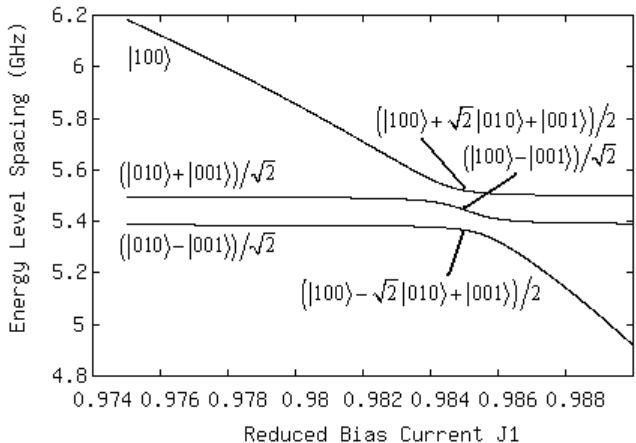

Fig. 4. Energy Level Spectrum for the linear three qubit system where $J_2$ = 0.98439, $J_3$ = 0.985, while $J_1$ is being ramped. When $J_1 = J_3 = 0.985$, the system forms a more symmetric set of entangled eigenstates. Towards the left of the figure, when $J_1$ is far from $J_2$ and $J_3$, qubits 2 and 3 become entangled with symmetric eigenstates.

Another property of the linear system is its ability to perform a $\pi$ shift in the phase between the two terms. Suppose the system is initialized in the symmetric eigenstate about qubits 2 and 3, with qubit 1 greatly detuned, as in Fig. 4. If $J_1$ is adiabatically increased through resonance with the other two qubits until qubit 1 is once again detuned, then the system will have switched to the anti-symmetric eigenstate about qubits 2 and 3, thus, performing a $\pi$ phase shift on the system while maintaining the original eigenstates.

### B. Triangular System

Another interesting qubit configuration is the triangular three-qubit system, as shown in Fig. 1b. The triangular system can be tuned to demonstrate most of the properties of its corresponding linear system. In particular, when $J_2$ = 0.979, $J_3$ = 0.985 and $J_1$ is ramped through these values, an energy level spectrum similar to that in Fig. 2 is produced. However, because each qubit is equally connected to each other in the triangular network, the energy gaps at the avoided crossings are the same.

Another feature of the system is shown in Fig. 5 where detuning one of the qubits results in a symmetric set of eigenstates. For instance, detuning $J_1$ results in symmetric and anti-symmetric states about qubits 2 and 3, without having to detune qubits 2 and 3. Thus, state transfer is possible between any of two qubits by detuning the third one.

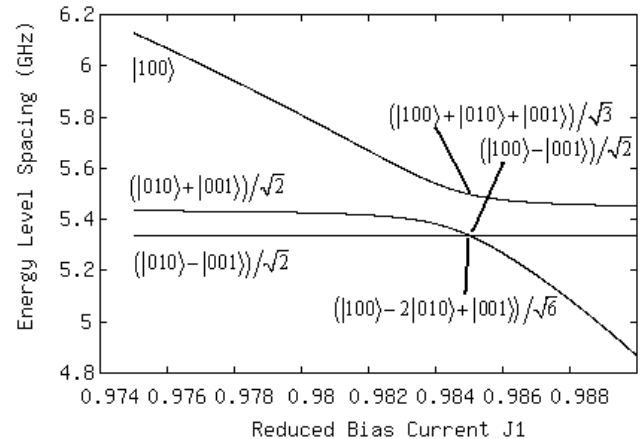

Fig. 5. Energy level spectrum for the triangular three qubit system. Here $J_2$ = 0.985, $J_3$ = 0.985, while $J_1$ is being ramped. When all three bias currents are equal a degeneracy is formed due to the symmetry of the system. The eigenstates at the degeneracy can actually be any linear combination of the two states indicated here. When $J_1$ is far from $J_2$ and $J_3$, qubits 1 and 3 become maximally entangled.

Of particular interest is the natural degeneracy that occurs when all three bias currents are equal. This is a result of the symmetry of the system where all qubits are connected equally to each other. If the bias currents are slightly off, the degeneracy is broken and a small gap between the two lowest energy levels appear.

### C. Four-qubit Systems

The Hamiltonian for a linear four-qubit system can be generated following a similar procedure. There is an asymmetry in the eigenstates that cannot be fixed by any slight detuning as was demonstrated in the linear three qubit system. This occurs because of the symmetry of the system. As a consequence of this asymmetry, perfect state transfer between qubits 1 and 4 is not possible. However, by significantly detuning either of the end qubits, the remaining three can be tuned to display the properties of the linear three qubit system. Thus state information can, in principle, be transferred from one end to the other, just not in one step.

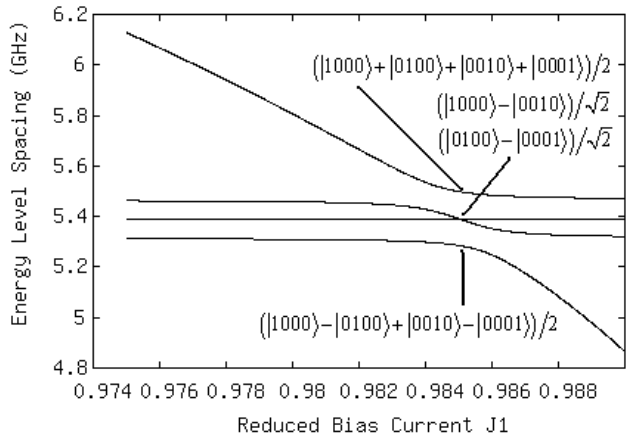

Fig. 6 Energy level spectrum for the four qubit box system. Here $J_2 = J_3 = J_4 = 0.985$, while $J_1$ is being ramped. When all four bias currents are equal the system forms eigenstates which exhibit a symmetry between qubits at opposite corners of the box.

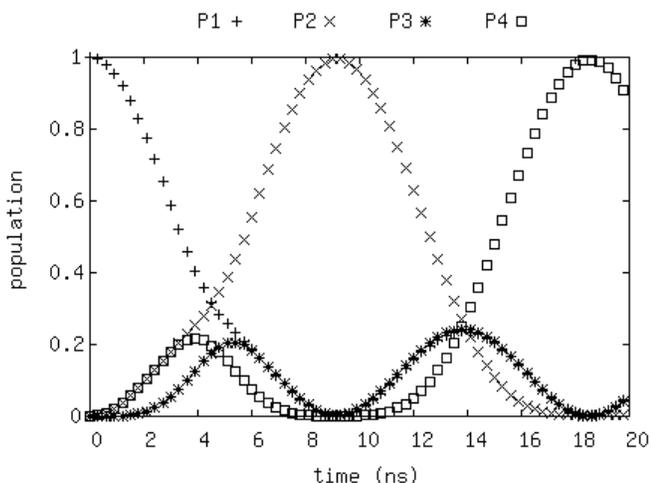

Fig. 7. The populations of the excited state being in qubits 1 through 4 of the four qubit box system are plotted. All four reduced bias currents are held at 0.985 until 2.7 ns have passed. Then $J_1$ and $J_2$ are changed to 0.9815 while the $J_3$ and $J_4$ stay the same. After 4 ns have passed $J_1$ and $J_2$ are changed back to 0.985 and the system evolves normally. This results in the excited state population being transferred to qubit 2.

The four qubit box graph, as seen in Fig. 1d, is the two-dimensional case of a general hypercube network of qubits [13], [15]. Like the linear four qubit system, it can be tuned to exhibit properties of three linear qubits. When one of the junctions is detuned from the others, the remaining three qubits emulate the properties and eigenstates of the slightly detuned three qubit system, similar to that demonstrated in Fig. 4. As can be seen in Fig. 6, when all the bias currents are the same, the eigenstates are symmetric and anti-symmetric about qubits on opposite corners of the four-qubit box. The anti-symmetric states are the two eigenstates at the degeneracy seen in Fig. 6, while the symmetric states are superpositions of the other two eigenstates. Adding the top and bottom states together produces a symmetric state about qubits 1 and 3; subtracting these two states produces a symmetric state about qubits 2 and 4.

This results in perfect state transfer between qubits on opposite corners of the box. It is possible to reroute the state information to any of the other qubits by momentarily detuning the starting qubit and the new target qubit. This detuning should take place when the state population is evenly distributed amongst the four qubits, and last just long enough for the relative phase of the qubit opposite the target qubit to increase by $\pi$ [17]. One example is to initialize the system with qubit 1 in its first excited state while the rest are in their ground states. Fig. 7 shows, as a function of time, all of the individual qubits' excited state populations. Left on its own, the excited state would go to qubit 3. In order to transfer the state information to junction 2, qubits 1 and 2 are detuned from the others at around 3 ns, but then brought back into resonance after a short time. Then the system naturally evolves until qubit 2 is in the excited state.

This method of momentarily detuning some of the qubits in the 2D hypercube to reroute quantum information can be applied to hypercube systems of higher dimensions. This allows for quantum state transfer between any of the nodes in a large network of qubits. The advantage of this method is that the amount of time the qubits need to be detuned is dependent on how much they are detuned. If the qubits are detuned for a longer period of time, then the detuning need not be as great. This allows for some flexibility when determining how the network will be programmed. If the hypercube is to be able to act as a network between different qubit systems, then it will need to be sufficiently detuned from those systems when it is transferring information. Being able to route information along that network without having to detune parts of it very much will help minimize the interaction with the rest of the system.

## IV. CONCLUSION

We have performed theoretical simulations of three and four capacitively-coupled phase qubits arranged in different network geometries. Avoided crossings in their energy level spectra, along with the eigenstates of such entangled systems were calculated. We have demonstrated how these eigenstates could be manipulated through slight detuning of one or more of the qubits. With the tunability of the Josephson phase qubits, we have the ability to change the eigenstates of these coupled systems in order to exhibit quantum state transfer between any of the qubits in the system.

## ACKNOWLEDGMENT

We would like to thank the Superconducting Quantum Computing Group at University of Maryland, College Park and Professor Robert Gilmore from the Department of Physics, Drexel University for helpful discussions.